\documentclass[times, twoside, watermark]{zHenriquesLab-StyleBioRxiv}
\usepackage{blindtext}

\leadauthor{Nierula} 

\begin{document}

\title{Differential Physiological Responses to Proxemic and Facial Threats in Virtual Avatar Interactions}
\shorttitle{Physiological Responses to Proxemic and Facial Threats}

\author[1,\Letter]{Birgit Nierula}
\author[1]{Mustafa Tevfik Lafci}
\author[1]{Anna Melnik}
\author[1]{Mert Akgül}
\author[1]{Farelle Toumaleu Siewe}
\author[1]{Sebastian Bosse}

\affil[1]{Fraunhofer Heinrich-Hertz-Institute, Vision and Imaging Technologies Department, Interactive \& Cognitive Systems Group, Einsteinufer 37, 10587 Berlin, Germany}

\maketitle

\begin{abstract}
Proxemics, the study of spatial behavior, is fundamental to social interaction and increasingly relevant for virtual reality (VR) applications. While previous research has established that users respond to personal space violations in VR similarly as in real-world settings, phase-specific physiological responses and the modulating effects of facial expressions remain understudied. We investigated physiological and subjective responses to personal space violations by virtual avatars, to understand how threatening facial expressions and interaction phases (approach vs. standing) influence these responses. Sixteen participants experienced a 2×2 factorial design manipulating Personal Space (intrusion vs. respect) and Facial Expression (neutral vs. angry) while we recorded skin conductance response (SCR), heart rate variability (HRV), and discomfort ratings. Personal space boundaries were individually calibrated using a stop-distance procedure. Results show that SCR responses are significantly higher during the standing phase compared to the approach phase when personal space was violated, indicating that prolonged proximity within personal space boundaries is more physiologically arousing than the approach itself. Angry facial expressions significantly reduced HRV, reflecting decreased parasympathetic activity, and increased discomfort ratings, but did not amplify SCR responses.  These findings demonstrate that different physiological modalities capture distinct aspects of proxemic responses: SCR primarily reflects spatial boundary violations, while HRV responds to facial threat cues. Our results provide insights for developing comprehensive multi-modal assessments of social behavior in virtual environments and inform the design of more realistic avatar interactions.
\end{abstract}


\begin{corrauthor}
birgit.nierula\at hhi.fraunhofer.de
\end{corrauthor}
\section*{Introduction} 
Proxemics refers to the study of how humans perceive and use space during communication and is a core pillar of non-verbal communication~\cite{hall_hidden_1966}. This field explores how spatial behavior reflects cultural context and is fundamental to organizing social interaction, making it particularly relevant for understanding avatar interactions in virtual environments. A key aspect of proxemics is the concept of personal space, which refers to the invisible space directly surrounding a person; violations to this space trigger physiological arousal and experienced discomfort~\cite{hayduk_personal_1978}. Personal space represents a dynamic perception that plays an important role in social communication and behavior across species~\cite{hayduk_1983, graziano_2006}. For example, the extent and shape of personal space boundaries can be influenced by personal factors~\cite{mello_navigating_2024}, social cues~\cite{ruggiero_effect_2017}, or cultural norms~\cite{iachini_near_2015}. \\
Regarding personal space perception, virtual reality (VR) users exhibit behaviors similar to those in real-world settings. Multiple studies confirm that users react to personal space violations in single- and multi-user VR scenarios~\cite{bailenson_interpersonal_2003, llobera_proxemics_2010, hecht_shape_2019, tootell_psychological_2021,yee_unbearable_2007}. Beyond reported discomfort, participants exhibit elevated skin conductance response (SCR) to personal space violations in VR, with magnitudes comparable to those observed during violations between real humans~\cite{tootell_psychological_2021}, and amplified responses with the number of avatars~\cite{llobera_proxemics_2010}. In the current study, we were interested in whether a threatening facial expression of the avatar would lead to additional changes in discomfort ratings and physiological parameters. Previous work demonstrates that angry facial expressions increase personal space~\cite{ruggiero_facialexp_2017b}. Furthermore, angry faces evoke stronger SCRs than neutral faces~\cite{clark_1992, juuse_facialexp_2024} and elicit stronger changes in heart rate than happy faces~\cite{critchley_facialexp_2005}. Previous studies either investigated the phase when personal space was intruded upon in isolation~\cite{tootell_psychological_2021} or examined approach and standing phases together~\cite{llobera_proxemics_2010}. However, the approach of another human could by itself constitute a threat. \\
Consequently, the field still lacks (i) phase-specific autonomic markers of proxemic violations and (ii) evidence that an avatar's facial affect modulates those markers. We address both gaps by combining SCR with heart rate variability (HRV) and orthogonally manipulating avatar facial expression (neutral vs. angry) during individually calibrated personal space intrusions.
We tested the following hypotheses: H1—SCR will be larger during standing compared to the approach phases; H2—SCR will be larger for personal space intrusions than for respected distances; H3—angry facial expressions will further amplify SCR; H4—HRV will be smaller for personal space intrusions; H5—angry facial expressions will further decrease HRV, reflecting reduced parasympathetic activity.

\section*{Methods}

\subsection*{Participants}
Twenty healthy volunteers (10 females) with normal or corrected-to-normal vision participated in the experiment. Four datasets were excluded due to acquisition problems, yielding 16 participants (mean age=26±5 years). This sample size matches prior proxemics studies~\cite{llobera_proxemics_2010,tootell_psychological_2021}. Participants were naïve to the research question, provided written informed consent, and received monetary compensation. The local ethics committee approved the experiment.

\subsection*{Virtual Reality Environment}
The virtual environment was developed in Unity (version 2020.3.20f1) and presented via a Meta Quest Pro HMD (106$^{\circ}$ field of view, 1800~$\times$~1920~pixels per eye, 90~Hz frame rate). An androgynous avatar from the Unity Asset Store was selected to minimize gender perception confounds. The virtual room featured a standard living room environment.

\subsection*{Physiological Recordings}
\subsubsection*{Electrocardiography}
ECG was recorded with passive Ag/AgCl electrodes (Easycap, Wörthsee, Germany) placed according to Einthoven lead I on the left (positive) and right (negative) wrists, with ground over the left ankle's lateral malleolus.
\subsubsection*{Skin Conductance Response} 
SCR was recorded with two sensors (Brain Products GmbH, Gilching, Germany) on the second phalanx of the second and third digits of the non-dominant hand.\\
Additional EEG (64 electrodes) and respiratory data were recorded, but are not reported here, as they remain part of ongoing analysis.\\
Data were acquired using wireless LiveAmp amplifiers (Brain Products GmbH, Gilching, Germany) with third-order sinc filtering (-3 dB at 131 Hz) and a 500 Hz sampling rate. Lab Streaming Layer (LSL, www.github.com/sccn/labstreaminglayer) ensured temporal synchronization. Offline analysis was performed in Python (version 3.10.14) using the MNE-Python toolbox~\cite{gramfort_meg_2013}.

\subsection*{Stop-distance procedure (SDP)}
The SDP~\cite{burgess_social_1980, kaitz_adult_2004, williams_personal_1971} was used to assess individual personal space boundaries and calibrate avatar walking distances. Participants said "stop" when the slowly approaching avatar reached their comfort threshold. This distance served as a reference (100\%) for experimental manipulations. SDP was measured five times (before each block, plus at the end of the experiment) to control for habituation.

\subsection*{Experimental Procedure}
\begin{figure}[!h]
\centering
  \includegraphics[width=0.75\linewidth]{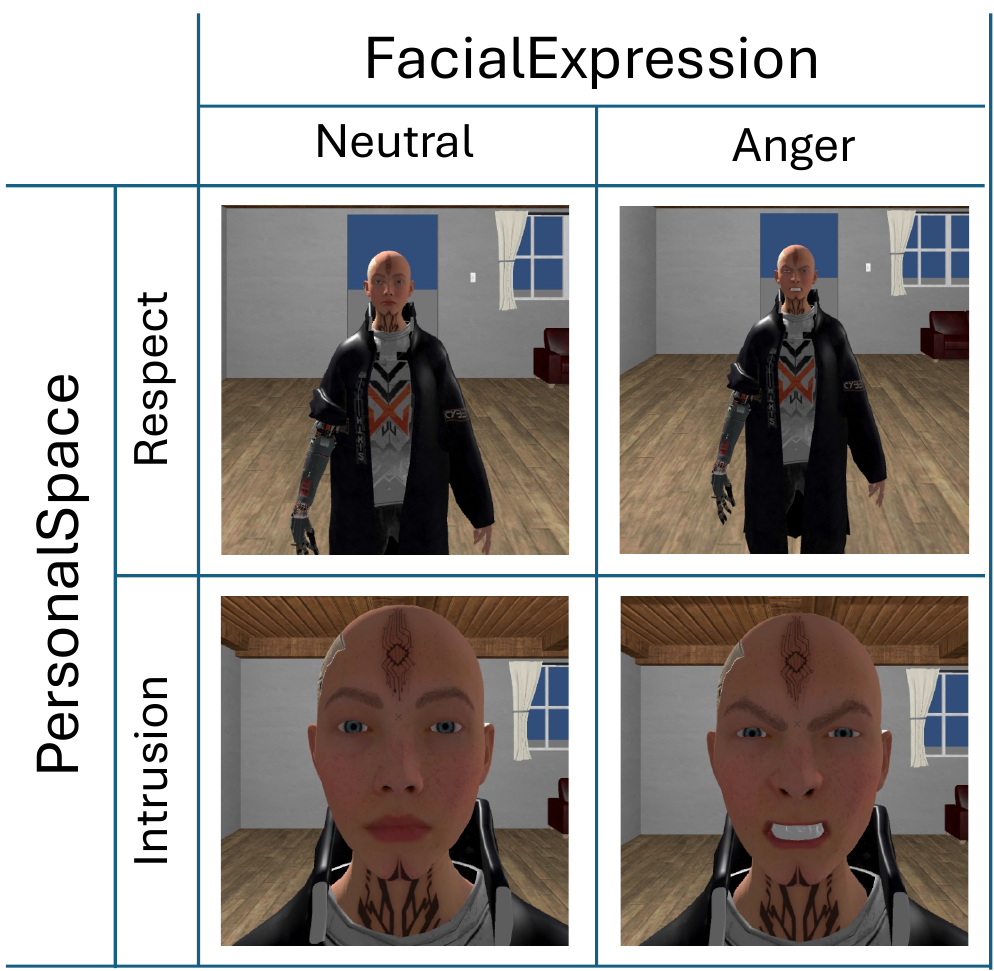}
  \caption{Experimental Conditions. The experiment employed a 2×2 factorial design with factors Personal Space (intrusion vs. respect) and  Facial Expression (neutral vs. angry), resulting in four conditions.}~\label{fig:figure1}
\end{figure}

The experiment used a 2×2 factorial within-subject design with factors Personal Space (intrusion vs. respect) and Facial Expression (neutral vs. angry). Figure~\ref{fig:figure1} displays the four experimental conditions (\textit {PSrespect+FEneutral}, \textit{PSrespect+FEangry}, \textit{PSintrusion+FEneutral}, and \textit{PSintrusion+FEangry}), which consisted of 15 trials randomly distributed across four blocks. 

\begin{figure}[!h]
\centering
  \includegraphics[width=\linewidth]{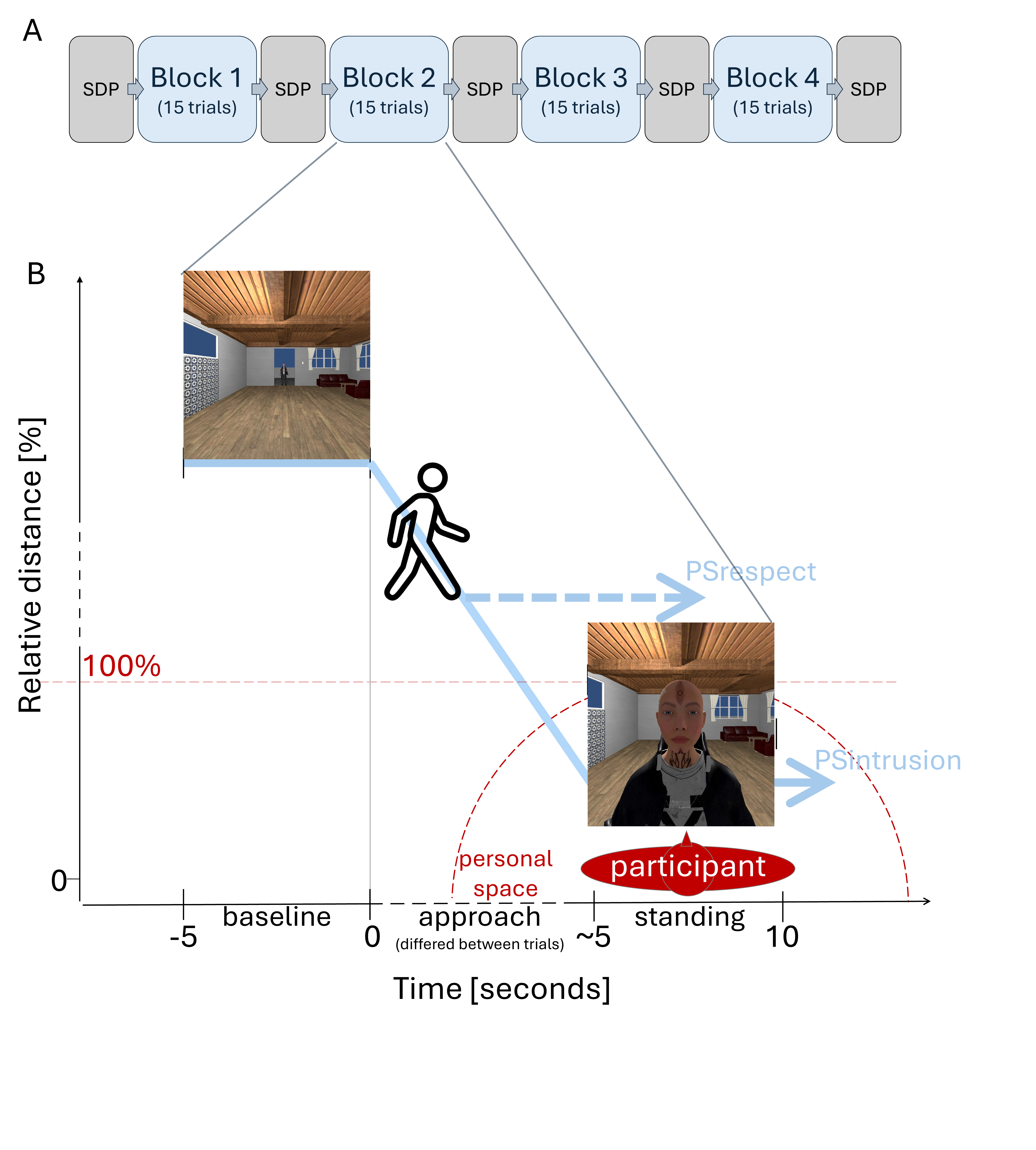}
  \caption{Experimental Design and Trial Structure. A. Four experimental blocks with stop-distance procedure (SDP) assessments before each block and after the final block. B. Trial timeline: Relative distance shows the distance between avatar and participant (100\% represents individual SDP). After a 5-second baseline (avatar at 8 m), the avatar approached with constant speed and remained standing for another 5-seconds. The avatar final positions either respected or intruded the personal space. Participants rated discomfort at the end of selected trials.}~\label{fig:figure2}
\end{figure}

Physiological sensors were placed while participants sat in an armchair. An experimenter then helped participants don the HMD; they stood up, and the avatar height was matched to their height. Next, participants were asked to explore and describe the virtual environment to enhance presence. Each block began with an SDP assessment followed by 15 avatar approach trials. Between blocks, participants were able to take a break and sit. A final SDP concluded the session (Figure~\ref{fig:figure2}A). Participants completed debriefing and avatar perception questionnaires.

Each trial began with the avatar standing for 5 seconds at 8 m distance, looking towards the participant (baseline phase). Participants were instructed to stand still throughout each trial and not to step backward or sideways. Next, the avatar started to walk at 1.2 m/s towards the participant (approach phase) and then stopped and stood for another 5 seconds in front of the participant (standing phase). The timeline of one trial is displayed in Figure~\ref{fig:figure2}B. The walking speed was kept constant throughout trials, and to avoid habituation, distances were jittered between 110\% - 170\% in the PSrespect conditions and between 28\% and 34\% in the PSintrusion conditions. These distances were based on the SDP measure directly before the experimental block and represented the personal space boundary set to 100\%. Experimental trials were presented in pseudo-randomized order such that in the course of all four blocks, participants went through 15 repetitions of each experimental condition.

\subsection*{Questionnaires}
After the 1st, the 7th, and the 15th (last) trial of the same condition, participants rated their level of discomfort on a 5-point Likert scale in the VR-environment.\\

\subsection*{Data Processing}
Recorded LSL streams from each block were transformed to MNE format. The raw files were cleaned of rectangular jump artifacts related to the wireless amplifiers.

\subsubsection*{SCR Preprocessing and Analysis}
SCR signal was bandpass filtered between 0.0159 Hz and 0.5 Hz (zero-phase IIR filter, Butterworth, first order)~\cite{bach_modelling_scr_2010}. The low-pass cut-off frequency was set at 0.5 Hz to eliminate high frequencies in the signal. Data underwent z-score transformation, and SCR peaks and amplitudes were identified using the NeuroKit2 toolbox~\cite{makowski_neurokit_2021}. Following manual trial inspection and noise rejection, maximal SCR amplitudes were extracted for the approach and standing phases. Amplitude values were averaged by participant, condition, and phase for statistical analysis. Four participants lacking SCR responses in at least one condition were excluded, yielding 12 subjects.  

\subsubsection*{ECG Preprocessing and Analysis}
ECG signals were z-score transformed, corrected for inversion, and filtered with a 0.5 Hz high-pass (Butterworth, 5th order) and 50 Hz notch filter. R-peaks were detected using NeuroKit2~\cite{makowski_neurokit_2021} and misdetections corrected using WFDB~\cite{xie2023wfdb} (available via PhysioNet~\cite{goldberger2000physiobank}) and visual inspection. Heart rate variability was quantified using the Root Mean Square of Successive Differences (RMSSD):
\begin{equation}
\text{RMSSD} = \sqrt{ \frac{1}{N - 1} \sum_{i=1}^{N-1} \left( RR_{i+1} - RR_i \right)^2 },
\end{equation}
where $RR_i$ is the $i^{th}$ R-R interval duration and $N$ the total intervals analyzed.

RMSSD was computed separately for baseline and standing phases. Since these time windows consisted of 5 seconds, we ensured that intervals consisted of at least four R-peaks (Mean: 7.24 and 7.08 peaks, respectively). While RMSSD is recommended for $\ge$10 seconds~\cite{salahuddin2007ultra, shaffer2020critical, krause_ushrv_2023}, phase-specific analysis was necessary given distinct trial phases. We hypothesized HRV effects would be most pronounced during standing, justifying shorter intervals. Two subjects with unrepairable signals were excluded, yielding 14 subjects. RMSSD values from the baseline and standing phases were used for statistical analysis.


\subsection*{Statistics}
All statistics were performed using RStudio (~\cite{rstudio_team_rstudio_2024}, R version 4.4.1). 
For the analysis of questionnaire data, we chose a Cumulative Link Mixed Effects Model (CLMM)  to appropriately handle the ordinal nature of our Likert scale data while accounting for the 2-factorial repeated measures design~\cite{taylor_rating_2022}. CLMM were performed using the \textit{ordinal} package in R (function: clmm)~\cite{rune_ordinal_2023}.
Physiological data was analyzed employing a Linear Mixed-effects Model (LMM) using the \textit{lme4} package~\cite{bates_fitting_2015} in R.
Significance values were set to $\alpha$ = 0.05.

\section*{Results}
\subsection*{Discomfort Ratings}
We modeled participants' discomfort employing a CLMM with the fixed effects of Personal Space (respected vs. intruded), Facial Expression (neutral vs. angry), and Repetition of the rating
(first vs. second vs. third). The model included a two-way interaction between Personal Space and Facial Expression, with Repetition entered as a main effect. A random intercept for each participant was specified to account for repeated measures within a subject.  

\begin{figure}[!h]
\centering
  \includegraphics[width=1\columnwidth]{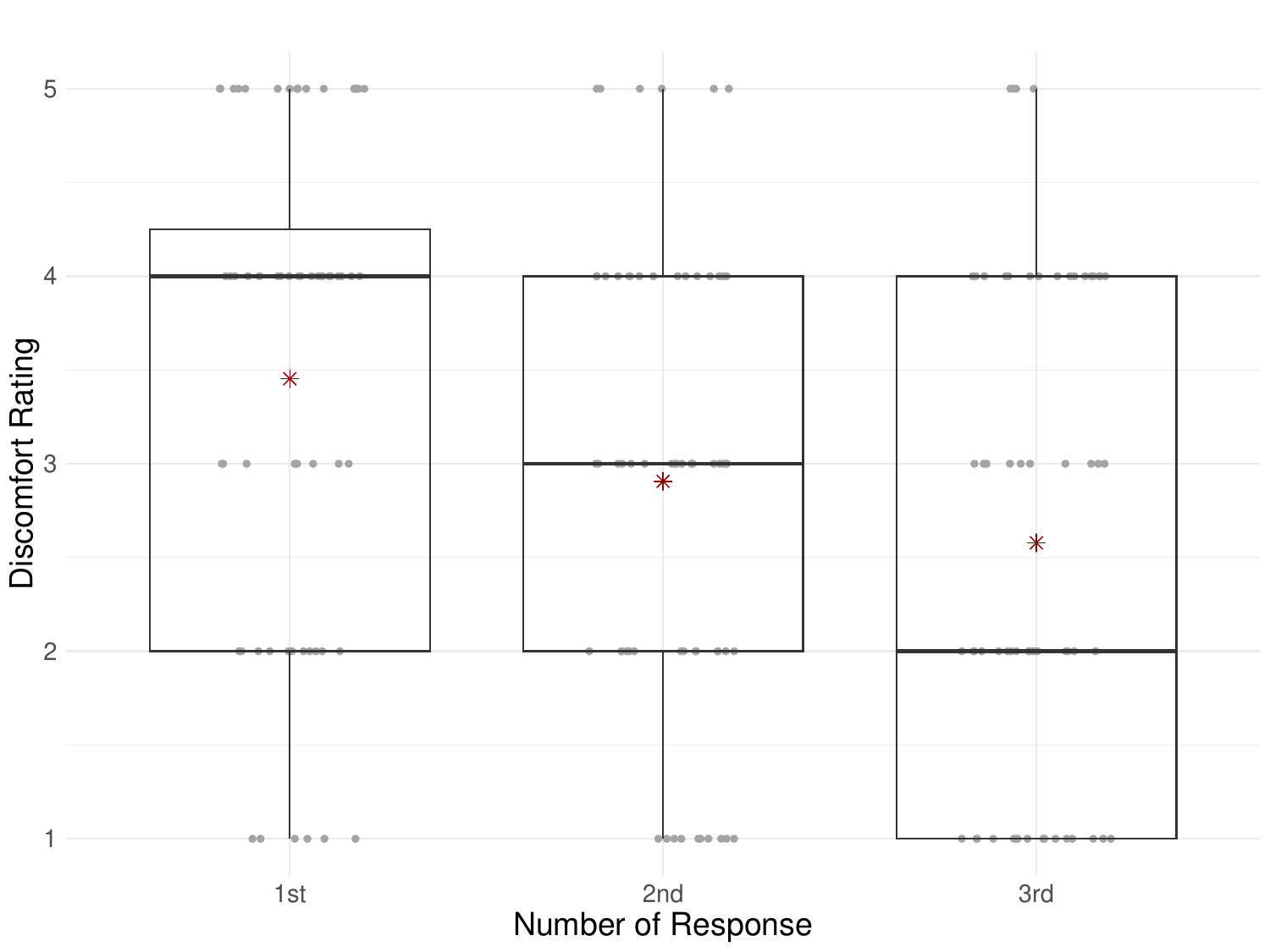}
  \caption{Boxplots of discomfort ratings over time. The median is displayed as horizontal line, the mean as red asterisk. Each condition was rated three times: after the 1st, the 7th, and the last (15th) trial. 
  }~\label{fig:figure3}
\end{figure}

We first looked at the changes in discomfort ratings over time, i.e., their first rating (after the 1st trial), their second rating (after the 7th trial), and their third rating (after the 15th trial) of the same condition. The CLMM model showed a significant linear fixed effect ($\beta$ = -1.360, \textit{SE} = 0.258, \textit{z} = -5.271, \textit{p} < 0.001), indicating that there were habituation effects in the discomfort ratings, which were fairly consistent from first to second and from second to third responses (see Figure~\ref{fig:figure3}).

There was a significant main effect of Personal Space ($\beta$ = 1.388, \textit{SE} = 0.217, \textit{z} = 6.392, \textit{p} < 0.001). Participants reported significantly higher discomfort when their personal space was intruded upon compared to when it was respected.

\begin{figure}[!h]
\centering
  \includegraphics[width=1\columnwidth]{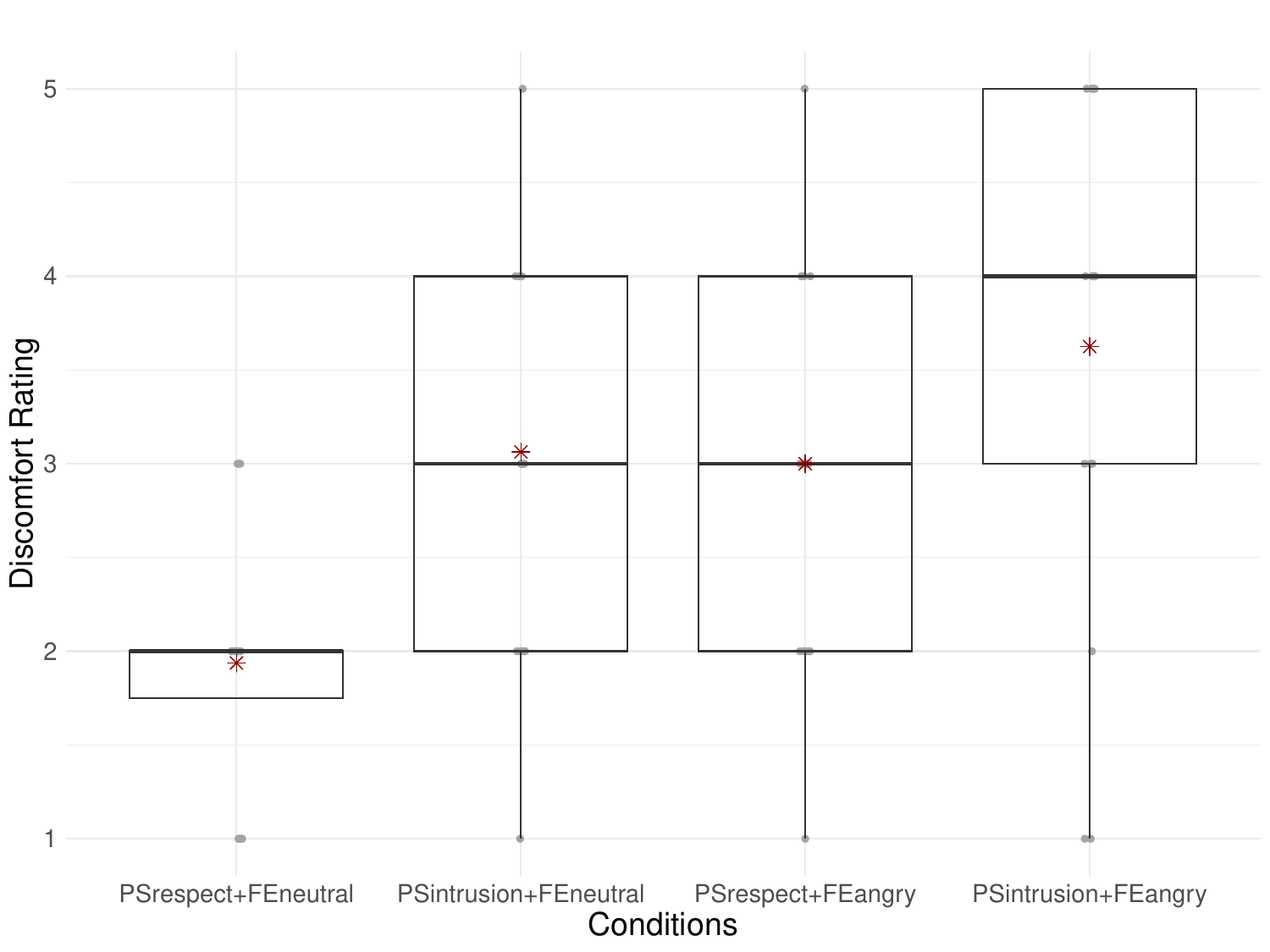}
  \caption{Boxplots of discomfort ratings. Horizontal lines displays the median, dark red asterisks the mean. 
  }~\label{fig:figure4}
\end{figure}

There was a significant main effect of avatar Facial Expression ($\beta$ = 1.163, \textit{SE} = 0.211, \textit{z} = 6.392, \textit{p} < 0.001). Participants reported significantly higher discomfort when the avatar displayed an angry face compared to a neutral expression. These main effects are displayed in Figure~\ref{fig:figure4}. There was no interaction between Personal Space and Facial Expression.

%
%
\subsection*{Personal Space Size}
One participant was identified as an outlier and was therefore removed from the analysis. Over all participants, individual personal space size was at Mean = 147 cm (\textit{SE} = 3 cm), there was no effect of gender (males: Mean = 140 cm, \textit{SE} = 4 cm; females: Mean = 151 cm, \textit{SE} = 5 cm; Welch Two Sample t-test: \textit{t}(12.9) = -0.737, \textit{p} = 0.474). \\
The 5 personal space assessments taken throughout the VR experience were analyzed using an LMM with the fixed effects of Time Factor (specified as five repeated assessments over the experiment) as a fixed effect. The model included a random intercept for each participant to account for repeated measures within a subject.

\begin{figure}[!h]
\centering
  \includegraphics[width=1\columnwidth]{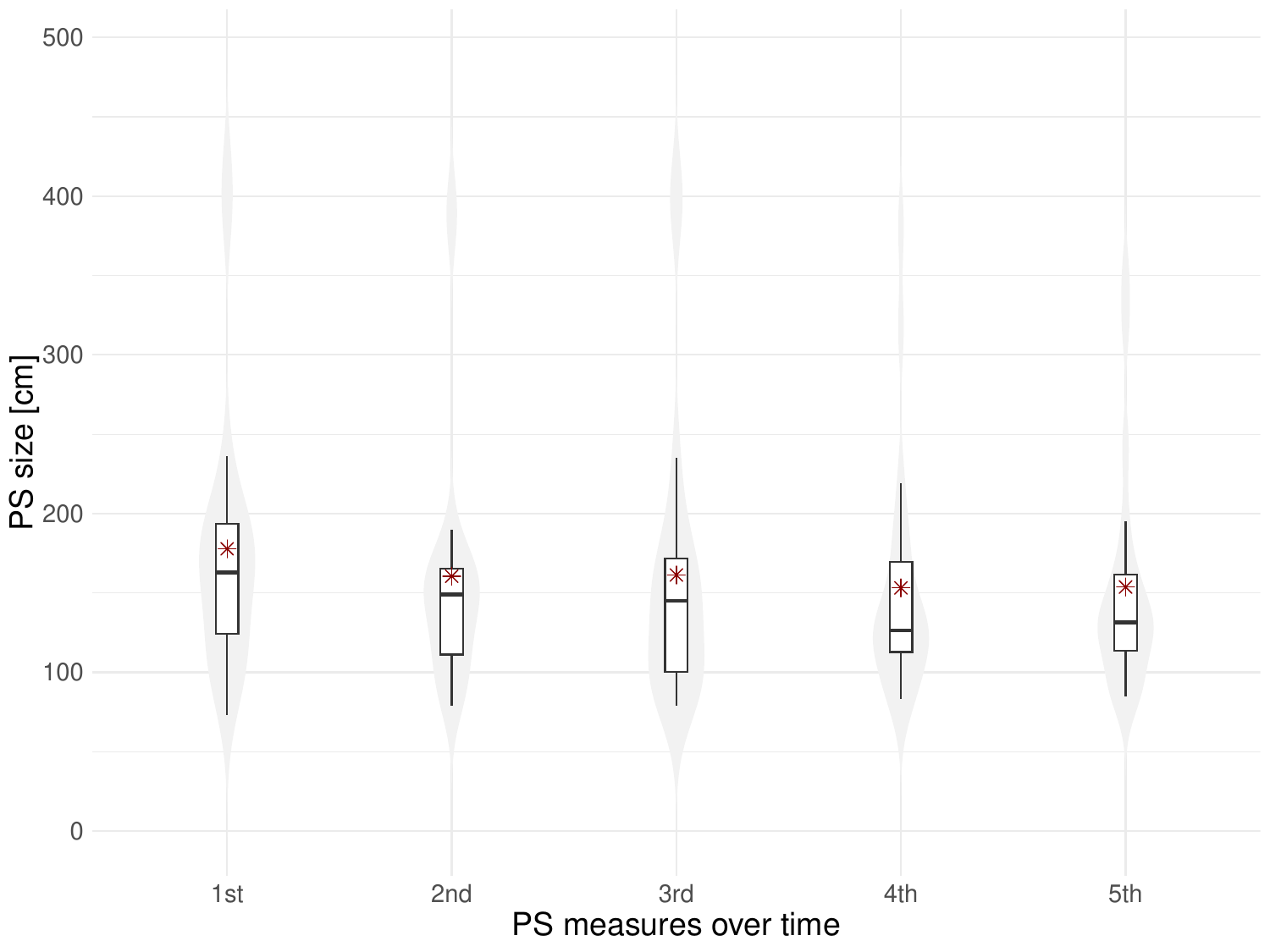}
  \caption{Personal Space size over time. Personal space measured before each block and after the final block using stop-distance procedure (SDP). Horizontal bold bars indicate the median, red asterisks -- the mean.
  }~\label{fig:figure5}
\end{figure}

There was a significant linear decrease in personal space size over time ($\beta$ = -14.8 cm, \textit{SE} = 6.2, \textit{t}(56) = -2.376, \textit{p} = 0.021). The baseline (intercept) personal space size was estimated at 146.6 cm (\textit{SE} = 7.6, \textit{t}(14) = 19.255, \textit{p} < 0.001). Figure~\ref{fig:figure5} displays a violin graph of personal space measures over the time course of the experiment (i.e., before each block and at the end of the last block).

\subsection*{Skin Conductance Response}
Participants who did not show SCR responses in one of the conditions were excluded from this analysis. To the remaining data of 12 participants, we fitted an LMM predicting SCR amplitude values with the fixed effects of Personal Space (respected vs. intruded), Facial Expression (neutral vs. angry), Phase (approach vs. stand), and all possible interactions between these three factors. The model included a random intercept for each participant to account for repeated measures within a subject.

\begin{figure}[!h]
\centering
  \includegraphics[width=1\columnwidth]{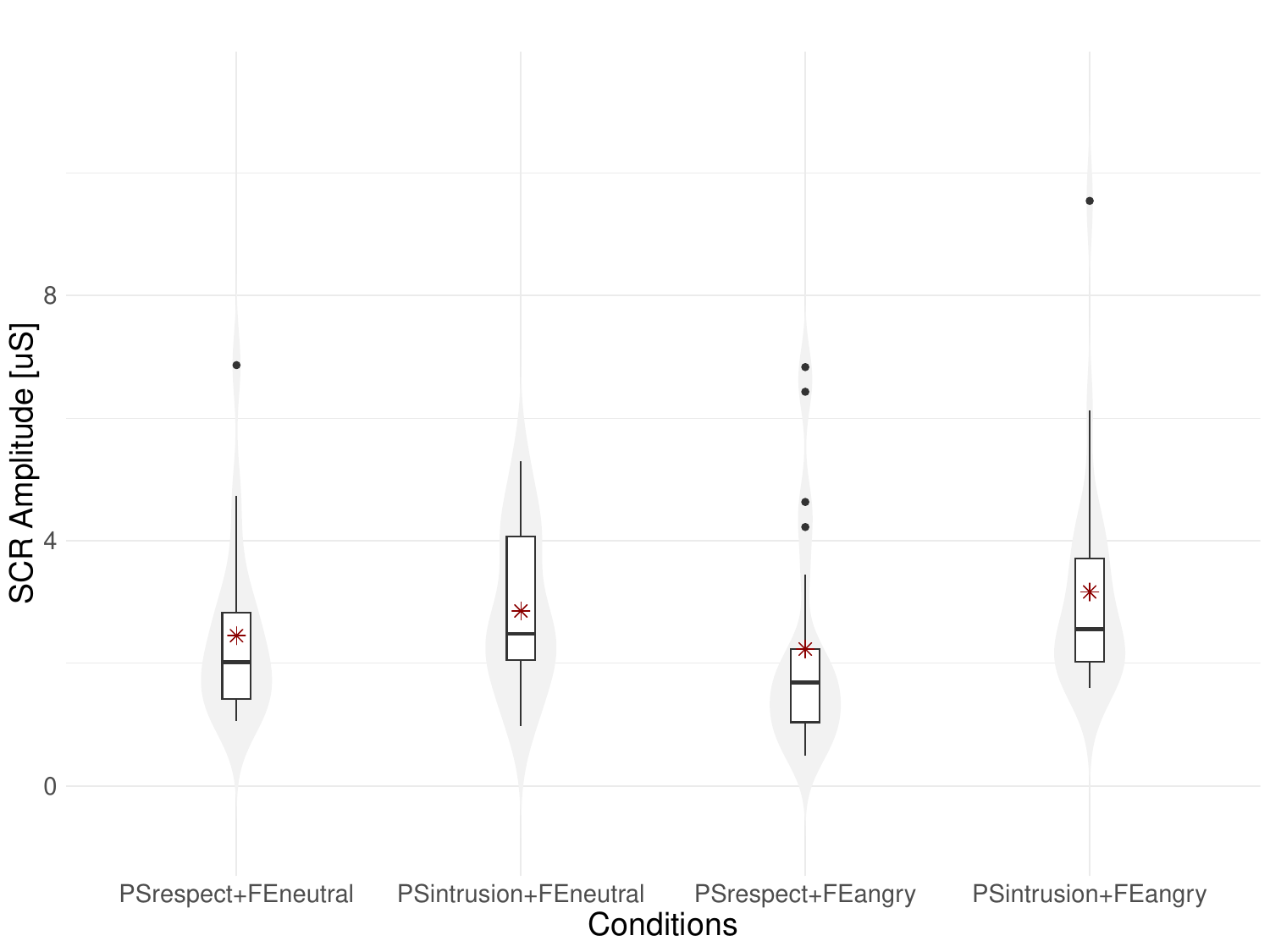}
  \caption{Skin conductance response (SCR) in the four different experimental conditions. The median is marked by the bold horizontal line and the mean by the red asterisk. Outliers beyond 1.5 times the inter-quartile range are plotted individually. 
  }~\label{fig:figure6}
\end{figure}

A Type III ANOVA test for fixed effects, which reflect the unique contribution of each factor after accounting for all other predictors in the model, revealed a significant main effect of Personal Space intrusion, \textit{F}(1, 65.272) = 4.760, \textit{p} = 0.033, indicating that intrusion into Personal Space reliably modulated physiological arousal. However, the corresponding fixed effect coefficient was not significant in the model summary ($\beta$ = -0.624, \textit{SE} = 0.668, \textit{t}(64.856) = -0.935, \textit{p} = 0.353, $\eta^2_p$ = 0.07), suggesting that this main effect is modulated by interaction terms with a medium effect size. Figure~\ref{fig:figure6} presents a violin plot showing the distribution and means of SCR amplitudes across experimental conditions.

No significant main effect of Facial Expression was observed ($\beta$ = -0.828, \textit{SE} = 0.668, \textit{t}(64.856) =  -1.240, \textit{p} = 0.220), suggesting that angry versus neutral avatar faces did not modulate SCR amplitudes overall. Similarly, the main effect of the trial Phase (approach vs. stand) was not statistically significant ($\beta$ = -1.284, \textit{SE} = 0.708, \textit{t}(67.067) = -1.813, \textit{p} = 0.074), although a trend toward reduced SCR in the standing phase compared to the approach phase was observed.

\begin{figure}[!h]
\centering
  \includegraphics[width=1\columnwidth]{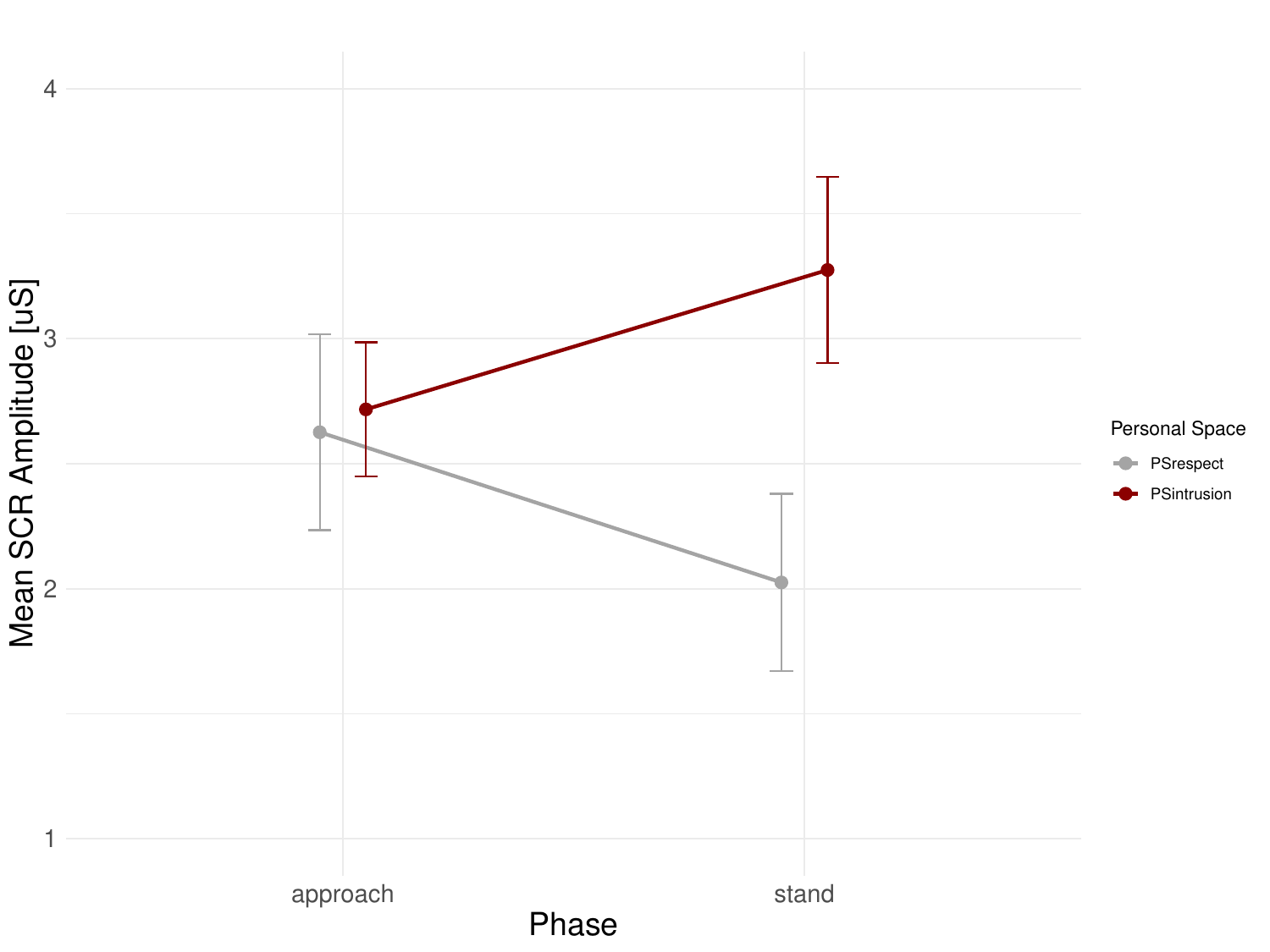}
  \caption{Interaction between Personal Space and Phase. Mean SCR amplitudes are plotted for the approach and standing trial phases for the two Personal Space conditions. Error bars represent standard errors. 
  }~\label{fig:figure7}
\end{figure}

Importantly, a significant interaction between Personal Space and trial Phase was found ($\beta$ = 2.075, \textit{SE} = 0.950, \textit{t}(66.147) = 2.184, \textit{p} = 0.033, $\eta^2_p$ = 0.05). This indicates that the effect of Personal Space intrusion on SCR amplitudes was specifically present in the standing phase, rather than during the approach. A post-hoc comparison within the standing phase revealed a substantial increase in SCR amplitude when the avatar intruded into personal space compared to when it did not (M = 3.30 vs. 1.98, Cohen’s \textit{d} = -0.89, 95\% CI [-1.54, -0.24]). In other words, SCR amplitudes reliably increased when the avatar remained within the participant's Personal Space boundary, confirming that SCR is a sensitive index of personal space violation at the expected time point (Figure~\ref{fig:figure7}). \\
No other interactions reached significance. The three-way interaction between Personal Space, Facial Expression, and Phase did not yield a reliable effect ($\beta$ = -1.754, \textit{SE} = 1.314, \textit{t}(65.543) = -1.334, \textit{p} = 0.187), suggesting that Facial Expression did not moderate the relationship between Personal Space and trial Phase in SCR responses.

\subsection*{Heart Rate Variability (HRV)}
Two participants' ECG data were excluded due to acquisition failure. From the remaining 14 participants, we analyzed RMSSD values using an LMM with the fixed effects of Personal Space (respected vs. intruded), Facial Expression (neutral vs. angry), and Phase (baseline vs. stand). The model included a two-way interaction between Personal Space and Phase, with Facial Expression entered as a main effect only. Random intercepts were modeled for each participant to account for repeated measures within a subject. 

The fixed effects analysis revealed a significant main effect of Facial Expression ($\beta$ = –3.866, \textit{SE} = 1.785, \textit{t}(94) = -2.166, \textit{p} = 0.033, $\eta^2_p$ = 0.05), indicating lower HRV (RMSSD) when participants observed an angry face compared to a neutral expression with a medium effect size. The Type III ANOVA confirmed the significant effect of Facial Expression (\textit{F}(1,94) = 4.692, \textit{p} = 0.033). These results indicate that threatening facial cues decreased HRV, consistent with autonomic nervous system activation in response to facial threat stimuli.

\begin{figure}[!h]
\centering
  \includegraphics[width=1\columnwidth]{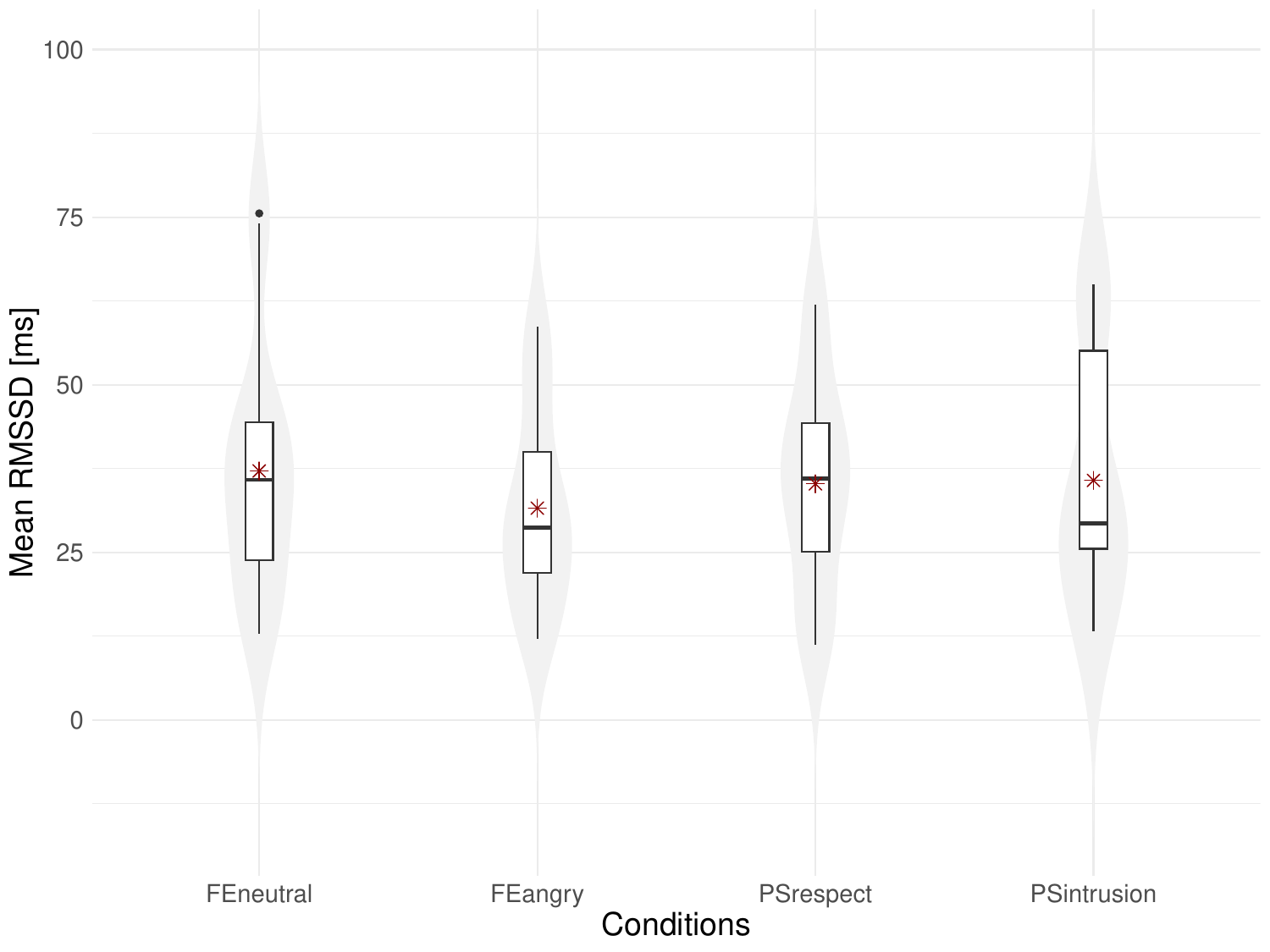}
  \caption{Heart Rate Variability (HRV) across Facial Expression and Personal Space conditions (median: horizontal bars, mean: red asterisks). 
  }~\label{fig:figure8}
\end{figure}

No significant main effects were found for Personal Space intrusion ($\beta$ = 2.202, \textit{SE} = 1.785, \textit{t}(94) = 1.234, \textit{p} = 0.220) or trial Phase ($\beta$ = –2.38, \textit{SE} = 1.785, \textit{t}(94) = –1.332, \textit{p} = 0.186). Figure~\ref{fig:figure8} shows a violin plot of RMSSD values summarized separately across Facial Expression and Personal Space conditions.

The interaction between Personal Space and Phase was also not significant ($\beta$ = 2.366, \textit{SE} = 2.524, \textit{t}(94) = 0.94, \textit{p} = 0.351). 

\section*{Discussion}
In the present study we aimed at advancing our understanding of proxemic behavior in virtual environments through multi-modal physiological assessment of our responses to personal space violations. We investigated whether threatening facial expressions modulate these responses and the effect of the approach itself.

Our findings demonstrate that arousal, measured via SCR amplitudes, is stronger when the avatar is standing in participants' personal space compared to when approaching (supporting H1). Our results further demonstrate stronger arousal (again measured via SCR amplitudes) and higher discomfort ratings for personal space violations in a phase-specific manner (partially supporting H2). Finally, our data shows a reduction of HRV with threatening facial cues (supporting H5). Our data did not support H3 and H4. In the following, we discuss these results in the context of the literature.

\subsection*{Personal Space Violation, Arousal and Discomfort}
Before entering a person's personal space the avatar has to approach the person, and this approach by itself could be perceived as a threat. Our data showed that it indeed made a difference for participant's arousal whether the avatar was in the approach or in the standing phase. Specifically, when the avatar was violating the participant's personal space, skin conductance responses were larger when standing directly in front of the participant compared to when approaching. These results are in line with previous findings in VR proxemics~\cite{tootell_psychological_2021, llobera_proxemics_2010}, which show increased SCR amplitudes to personal space violation. Our study contributes insight into phase-specific modulations of arousal by separately looking at the approach and standing phases. This was supported by discomfort ratings participants gave after 20\% of trials, which show stronger discomfort during violations of personal space. The skin conductance response is believed to be purely produced by the sympathetic part of the autonomous nervous system~\cite{critchleyReviewElectrodermalResponses2002}. Increased SCR amplitudes can be interpreted as a measure of arousal, with larger amplitudes indicating greater emotional or physiological arousal~\cite{dawsonElectrodermalSystem2016}. HRV was not modulated by violations of personal space. 

\subsection*{Facial Threat Cues}
An approaching avatar with an angry facial expression should be perceived as a stronger threat than one with a neutral expression. Our data support this on a physiological level, showing that facial threat cues reduced ultra-short HRV. This is believed to reflect decreased parasympathetic activity: Parasympathetic nerves, such as the vagal nerve, exert their effects more rapidly (<1s) than sympathetic nerves (>5s)~\cite{nunan_hrv_2010}, which is why variations in beat-to-beat intervals are considered a reflection of vagal nerve outflow~\cite{bigger_1989, kleiger_2005} and ultra-short HRV has thus been linked to parasympathetic activity. Several studies show that RMSSD is a valid measure for ultra-short HRV~\cite{thongAccuracyUltrashortHeart17, nussinovitchReliabilityUltraShortECG2011, munozValidityUltraShortRecordings2015}, especially when taken as average over several trials~\cite {munozValidityUltraShortRecordings2015}. Although our 5s windows fall below the recommended 10s~\cite{thongAccuracyUltrashortHeart17, nussinovitchReliabilityUltraShortECG2011, munozValidityUltraShortRecordings2015}, we believe our HRV results are valid because i) parasympathetic reaction can be expected within 1s~\cite{nunan_hrv_2010}, ii) our RMSSD measures are averaged over 15 trials and are thus less prone to biases~\cite {munozValidityUltraShortRecordings2015}, iii) we ensured single RMSSD measures were based on 4 or more peak-to-peak intervals (on average 7), and iv) provide thus the first evidence for HRV measures in such short time windows. We acknowledge ongoing debates about ultra-short HRV reliability~\cite{burma_2021}, particularly for clinical applications where measurement errors affect diagnostics, and believe that our reported HRV results are useful for the affective computing field. The reported reduction in HRV to facial threat cues is in line with related work showing reduced HRV to fear~\cite{wuHowAmusementAnger2019} or unpleasant emotional facial expressions~\cite{chungChangesHeartRate2025} and is supported by increased discomfort ratings during the angry facial expression condition in our study as well as in previous work~\cite{ruggiero_facialexp_2017b}.  \\
SCR peak amplitudes were not sensitive to this change. In fact, we expected highest SCR peak amplitudes during personal space violation with facial threat cues, similar to violations of personal space boundaries with multiple avatars~\cite{llobera_proxemics_2010}.

\subsection*{Habituation Within a Single Session}
Over the course of the experimental sessions, participants' comfort distance--operationalized as personal space size and measured using the stop-distance procedure--decreased, indicating that participants habituated to the avatar's presence and proximity. This effect was further supported by the concurrent reduction in subjective discomfort ratings across experimental blocks. To account for these temporal effects, our experimental design assessed personal space size and recalibrated the reference distance (100\% line in Figure~\ref{fig:figure2}) at the onset of each block.

\subsection*{Limitations and Future Work}
Several limitations should be acknowledged. First, our sample size was relatively small (n = 16), which may limit generalizability. Second, we used a single androgynous avatar, which may not represent the full range of social agents in VR environments. Third, we did not control for participants' prior VR experience or cultural background, both of which may influence proxemic behavior. Fourth, we limited the avatar's facial expression to neutral and angry. Future studies should employ larger samples and examine how avatar characteristics (gender, ethnicity) and different facial expressions affect proxemic responses and how they interact with gender, ethnicity, and cultural background of the VR user. 

\section*{Conclusion}
We investigated the effects of personal space violations and facial threat cues on discomfort ratings, skin conductance responses (SCR), and heart rate variability (HRV) in a VR human-avatar interaction. While discomfort ratings reflected both manipulations, physiological measures showed distinct patterns: personal space violations selectively affected SCR, whereas facial threat cues selectively influenced HRV. These findings demonstrate that different physiological modalities capture distinct aspects of proxemic responses, providing insights for developing comprehensive multi-modal assessments of social behavior in virtual environments.


\begin{acknowledgements}
We would like to thank the AspektEins GmbH for their support with the angry face animation and Dilara Damar for her support during data collection. This research was funded by the German Ministry of Education and Research (K3VR, 13N16388), the Max Planck Society, and the Fraunhofer Society (project NeuroHum).
\end{acknowledgements}

\section*{Bibliography}
\bibliography{preprint}

\begin{thebibliography}{42}
\providecommand{\natexlab}[1]{#1}
\providecommand{\url}[1]{\texttt{#1}}
\expandafter\ifx\csname urlstyle\endcsname\relax
  \providecommand{\doi}[1]{doi: #1}\else
  \providecommand{\doi}{doi: \begingroup \urlstyle{rm}\Url}\fi

\bibitem[Hall(1966)]{hall_hidden_1966}
Edward~T. Hall.
\newblock \emph{The hidden dimension}.
\newblock Doubleday, 1966.

\bibitem[Hayduk(1978)]{hayduk_personal_1978}
Leslie~Alec Hayduk.
\newblock Personal space: An evaluative and orienting overview.
\newblock \emph{Psychol Bull}, 85\penalty0 (1):\penalty0 117--134, 1978.

\bibitem[Hayduk(1983)]{hayduk_1983}
Leslie~A Hayduk.
\newblock Personal space: where we now stand.
\newblock \emph{Psychol Bull}, 94\penalty0 (2):\penalty0 293, 1983.

\bibitem[Graziano and Cooke(2006)]{graziano_2006}
Michael~S.A. Graziano and Dylan~F. Cooke.
\newblock Parieto-frontal interactions, personal space, and defensive behavior.
\newblock \emph{Neuropsychologia}, 44\penalty0 (6):\penalty0 845--859, 2006.
\newblock \doi{10.1016/j.neuropsychologia.2005.09.009}.

\bibitem[Mello et~al.(2024)Mello, Welsch, Verbokkem, Knierim, and Dechant]{mello_navigating_2024}
Beatriz Mello, Robin Welsch, Marissa~Christien Verbokkem, Pascal Knierim, and Martin~Johannes Dechant.
\newblock Navigating the virtual gaze.
\newblock In \emph{Proceedings of the {CHI} Conference on Human Factors in Computing Systems}, pages 1--15, 2024.
\newblock \doi{10.1145/3613904.3642359}.

\bibitem[Ruggiero et~al.(2017)Ruggiero, Frassinetti, Coello, Rapuano, Di~Cola, and Iachini]{ruggiero_effect_2017}
Gennaro Ruggiero, Francesca Frassinetti, Yann Coello, Mariachiara Rapuano, Armando~Schiano Di~Cola, and Tina Iachini.
\newblock The effect of facial expressions on peripersonal and interpersonal spaces.
\newblock \emph{Psychol Res}, 81\penalty0 (6):\penalty0 1232--1240, 2017.
\newblock \doi{10.1007/s00426-016-0806-x}.

\bibitem[Iachini et~al.(2015)Iachini, Pagliaro, and Ruggiero]{iachini_near_2015}
Tina Iachini, Stefano Pagliaro, and Gennaro Ruggiero.
\newblock Near or far? it depends on my impression.
\newblock \emph{Acta Psychol}, 161:\penalty0 131--136, 2015.
\newblock \doi{10.1016/j.actpsy.2015.09.003}.

\bibitem[Bailenson et~al.(2003)Bailenson, Blascovich, Beall, and Loomis]{bailenson_interpersonal_2003}
Jeremy~N. Bailenson, Jim Blascovich, Andrew~C. Beall, and Jack~M. Loomis.
\newblock Interpersonal distance in immersive virtual environments.
\newblock \emph{Pers Soc Psychol Bull}, 29\penalty0 (7):\penalty0 819--833, 2003.
\newblock \doi{10.1177/0146167203029007002}.

\bibitem[Llobera et~al.(2010)Llobera, Spanlang, Ruffini, and Slater]{llobera_proxemics_2010}
Joan Llobera, Bernhard Spanlang, Giulio Ruffini, and Mel Slater.
\newblock Proxemics with multiple dynamic characters in an immersive virtual environment.
\newblock \emph{ACM Trans Appl Percept}, 8\penalty0 (1):\penalty0 1--12, 2010.
\newblock \doi{10.1145/1857893.1857896}.

\bibitem[Hecht et~al.(2019)Hecht, Welsch, Viehoff, and Longo]{hecht_shape_2019}
Heiko Hecht, Robin Welsch, Jana Viehoff, and Matthew~R. Longo.
\newblock The shape of personal space.
\newblock \emph{Acta Psychol}, 193:\penalty0 113--122, 2019.
\newblock \doi{10.1016/j.actpsy.2018.12.009}.

\bibitem[Tootell et~al.(2021)Tootell, Zapetis, Babadi, Nasiriavanaki, Hughes, Mueser, Otto, Pace-Schott, and Holt]{tootell_psychological_2021}
Roger B.~H. Tootell, Sarah~L. Zapetis, Baktash Babadi, Zahra Nasiriavanaki, Dylan~E. Hughes, Kim Mueser, Michael Otto, Ed~Pace-Schott, and Daphne~J. Holt.
\newblock Psychological and physiological evidence for an initial ‘rough sketch’ calculation of personal space.
\newblock \emph{Sci Rep}, 11\penalty0 (1):\penalty0 20960, 2021.
\newblock \doi{10.1038/s41598-021-99578-1}.

\bibitem[Yee et~al.(2007)Yee, Bailenson, Urbanek, Chang, and Merget]{yee_unbearable_2007}
Nick Yee, Jeremy~N. Bailenson, Mark Urbanek, Francis Chang, and Dan Merget.
\newblock The unbearable likeness of being digital: The persistence of nonverbal social norms in online virtual environments.
\newblock \emph{Cyberpsychol Behav}, 10\penalty0 (1):\penalty0 115--121, 2007.
\newblock \doi{10.1089/cpb.2006.9984}.

\bibitem[Ruggiero(2017)]{ruggiero_facialexp_2017b}
Gennaro Ruggiero.
\newblock The effect of facial expressions on peripersonal and interpersonal spaces.
\newblock \emph{Psychol Res}, 2017.

\bibitem[Clark et~al.(1992)Clark, Siddle, and Bond]{clark_1992}
Belinda~M. Clark, David~A.T. Siddle, and Nigel~W. Bond.
\newblock Effects of social anxiety and facial expression on habituation of the electrodermal orienting response.
\newblock \emph{Biol Psychol}, 33\penalty0 (2-3):\penalty0 211--223, 1992.
\newblock \doi{10.1016/0301-0511(92)90033-q}.

\bibitem[Juuse et~al.(2024)Juuse, Tamm, Lõo, Allik, and Kreegipuu]{juuse_facialexp_2024}
Liina Juuse, Diina Tamm, Kaidi Lõo, Jüri Allik, and Kairi Kreegipuu.
\newblock Skin conductance response and habituation to emotional facial expressions and words.
\newblock \emph{Acta Psychologica}, 251:\penalty0 104573, 2024.
\newblock \doi{10.1016/j.actpsy.2024.104573}.

\bibitem[Critchley et~al.(2005)Critchley, Rotshtein, Nagai, O'Doherty, Mathias, and Dolan]{critchley_facialexp_2005}
Hugo~D. Critchley, Pia Rotshtein, Yoko Nagai, John O'Doherty, Christopher~J. Mathias, and Raymond~J. Dolan.
\newblock Activity in the human brain predicting differential heart rate responses to emotional facial expressions.
\newblock \emph{NeuroImage}, 24\penalty0 (3):\penalty0 751--762, 2005.
\newblock \doi{10.1016/j.neuroimage.2004.10.013}.

\bibitem[Gramfort(2013)]{gramfort_meg_2013}
Alexandre Gramfort.
\newblock {MEG} and {EEG} data analysis with {MNE}-python.
\newblock \emph{Front Neurosci}, 7, 2013.
\newblock \doi{10.3389/fnins.2013.00267}.

\bibitem[Burgess(1980)]{burgess_social_1980}
J.~Wesley Burgess.
\newblock Social group spacing of rhesus macaque troops (macaca mulatta) in outdoor enclosures: Environmental effects.
\newblock \emph{Behav and Neural Biol}, 30\penalty0 (1):\penalty0 49--55, 1980.
\newblock \doi{10.1016/S0163-1047(80)90869-9}.

\bibitem[Kaitz et~al.(2004)Kaitz, Bar-Haim, Lehrer, and Grossman]{kaitz_adult_2004}
Marsha Kaitz, Yair Bar-Haim, Melissa Lehrer, and Ephraim Grossman.
\newblock Adult attachment style and interpersonal distance.
\newblock \emph{Attach Hum Dev}, 6\penalty0 (3):\penalty0 285--304, 2004.
\newblock \doi{10.1080/14616730412331281520}.

\bibitem[Williams(1971)]{williams_personal_1971}
John~L. Williams.
\newblock Personal space and its relation to extraversion-introversion.
\newblock \emph{Can J Behav Sci}, 3\penalty0 (2):\penalty0 156--160, 1971.
\newblock \doi{10.1037/h0082257}.

\bibitem[Bach et~al.(2010)Bach, Flandin, Friston, and Dolan]{bach_modelling_scr_2010}
Dominik~R. Bach, Guillaume Flandin, Karl~J. Friston, and Raymond~J. Dolan.
\newblock Modelling event-related skin conductance responses.
\newblock \emph{Int J Psychophysiol}, 75\penalty0 (3):\penalty0 349--356, 2010.
\newblock \doi{10.1016/j.ijpsycho.2010.01.005}.

\bibitem[Makowski et~al.(2021)Makowski, Pham, Lau, Brammer, Lespinasse, Pham, Schölzel, and Chen]{makowski_neurokit_2021}
Dominique Makowski, Tam Pham, Zen~J. Lau, Jan~C. Brammer, Fran{\c{c}}ois Lespinasse, Hung Pham, Christopher Schölzel, and S.~H.~Annabel Chen.
\newblock {NeuroKit}2: A python toolbox for neurophysiological signal processing.
\newblock \emph{Behav Res Methods}, 53\penalty0 (4):\penalty0 1689--1696, feb 2021.
\newblock \doi{10.3758/s13428-020-01516-y}.

\bibitem[Xie et~al.(2023)Xie, McCullum, Johnson, Pollard, Gow, and Moody]{xie2023wfdb}
Chen Xie, Lucas McCullum, Alistair Johnson, Tom Pollard, Brian Gow, and Benjamin Moody.
\newblock {Waveform Database Software Package (WFDB) for Python} (version 4.1.0).
\newblock \url{https://doi.org/10.13026/9njx-6322}, 2023.

\bibitem[Goldberger et~al.(2000)Goldberger, Amaral, Glass, Hausdorff, Ivanov, Mark, Mietus, Moody, Peng, and Stanley]{goldberger2000physiobank}
Ary~L Goldberger, Luis~AN Amaral, Leon Glass, Jeffrey~M Hausdorff, Plamen~Ch Ivanov, Roger~G Mark, Joseph~E Mietus, George~B Moody, Chung-Kang Peng, and H~Eugene Stanley.
\newblock Physiobank, physiotoolkit, and physionet: components of a new research resource for complex physiologic signals.
\newblock \emph{Circulation}, 101\penalty0 (23):\penalty0 e215--e220, 2000.

\bibitem[Salahuddin et~al.(2007)Salahuddin, Cho, Jeong, and Kim]{salahuddin2007ultra}
Lizawati Salahuddin, Jaegeol Cho, Myeong~Gi Jeong, and Desok Kim.
\newblock Ultra short term analysis of heart rate variability for monitoring mental stress in mobile settings.
\newblock In \emph{2007 29th annual international conference of the ieee engineering in medicine and biology society}, pages 4656--4659, 2007.

\bibitem[Shaffer et~al.(2020)Shaffer, Meehan, and Zerr]{shaffer2020critical}
Fred Shaffer, Zachary~M Meehan, and Christopher~L Zerr.
\newblock A critical review of ultra-short-term heart rate variability norms research.
\newblock \emph{Front neurosci}, 14:\penalty0 594880, 2020.

\bibitem[Krause et~al.(2023)Krause, Vollmer, Wittfeld, Weihs, Frenzel, D{\"o}rr, Kaderali, Felix, Stubbe, Ewert, V{\"o}lzke, and Grabe]{krause_ushrv_2023}
Elischa Krause, Marcus Vollmer, Katharina Wittfeld, Antoine Weihs, Stefan Frenzel, Marcus D{\"o}rr, Lars Kaderali, Stephan~B. Felix, Beate Stubbe, Ralf Ewert, Henry V{\"o}lzke, and Hans~J. Grabe.
\newblock Evaluating heart rate variability with 10\>second multichannel electrocardiograms in a large population-based sample.
\newblock \emph{Front Cardiovasc Med}, 10, 2023.
\newblock ISSN 2297-055X.
\newblock \doi{10.3389/fcvm.2023.1144191}.

\bibitem[Team(2024)]{rstudio_team_rstudio_2024}
{RStudio} Team.
\newblock {RStudio}: Integrated development environment for r, 2024.

\bibitem[Taylor et~al.(2022)Taylor, Rousselet, Scheepers, and Sereno]{taylor_rating_2022}
Jack~E. Taylor, Guillaume~A. Rousselet, Christoph Scheepers, and Sara~C. Sereno.
\newblock Rating norms should be calculated from cumulative link mixed effects models.
\newblock \emph{Behav Res}, 55\penalty0 (5):\penalty0 2175--2196, 2022.
\newblock ISSN 1554-3528.
\newblock \doi{10.3758/s13428-022-01814-7}.

\bibitem[Christensen(2023)]{rune_ordinal_2023}
Rune H.~B. Christensen.
\newblock \emph{ordinal---Regression Models for Ordinal Data}, 2023.

\bibitem[Bates et~al.(2015)Bates, Mächler, Bolker, and Walker]{bates_fitting_2015}
Douglas Bates, Martin Mächler, Ben Bolker, and Steve Walker.
\newblock Fitting linear mixed-effects models using \textbf{lme4}.
\newblock \emph{J Stat Soft}, 67\penalty0 (1), 2015.
\newblock \doi{10.18637/jss.v067.i01}.

\bibitem[Critchley(2002)]{critchleyReviewElectrodermalResponses2002}
Hugo~D. Critchley.
\newblock Review: {Electrodermal} {Responses}: {What} {Happens} in the {Brain}.
\newblock \emph{The Neuroscientist}, 8\penalty0 (2):\penalty0 132--142, 2002.
\newblock \doi{10.1177/107385840200800209}.

\bibitem[Dawson et~al.(2016)Dawson, Schell, and Filion]{dawsonElectrodermalSystem2016}
Michael~E. Dawson, Anne~M. Schell, and Diane~L. Filion.
\newblock The {Electrodermal} {System}.
\newblock In John~T. Cacioppo, Louis~G. Tassinary, and Gary~G. Berntson, editors, \emph{Handbook of Psychophysiology}, pages 217--243. Cambridge University Press, 4 edition, 2016.
\newblock ISBN 978-1-107-41578-2 978-1-107-05852-1.
\newblock \doi{10.1017/9781107415782.010}.

\bibitem[Nunan et~al.(2010)Nunan, Sandercock, and Brodie]{nunan_hrv_2010}
David Nunan, Gavin R.~H. Sandercock, and David~A. Brodie.
\newblock A {{Quantitative Systematic Review}} of {{Normal Values}} for {{Short-Term Heart Rate Variability}} in {{Healthy Adults}}.
\newblock \emph{Pacing Clin Electrophysiol}, 33\penalty0 (11):\penalty0 1407--1417, 2010.
\newblock \doi{10.1111/j.1540-8159.2010.02841.x}.

\bibitem[Bigger et~al.(1989)Bigger, Albrecht, Steinman, Rolnitzky, Fleiss, and Cohen]{bigger_1989}
J.~Thomas Bigger, Paul Albrecht, Richard~C. Steinman, Linda~M. Rolnitzky, Joseph~L. Fleiss, and Richard~J. Cohen.
\newblock Comparison of time- and frequency domain-based measures of cardiac parasympathetic activity in {{Holter}} recordings after myocardial infarction.
\newblock \emph{Am J Cardiol}, 64\penalty0 (8):\penalty0 536--538, 1989.
\newblock \doi{10.1016/0002-9149(89)90436-0}.

\bibitem[Kleiger et~al.(2005)Kleiger, Stein, and J~Thomas~Bigger]{kleiger_2005}
Robert~E. Kleiger, Phyllis~K. Stein, and Jr~J~Thomas~Bigger.
\newblock Heart {{Rate Variability}}: {{Measurement}} and {{Clinical Utility}}.
\newblock \emph{Ann Noninvas Electrocardiol}, 10\penalty0 (1):\penalty0 88, 2005.
\newblock \doi{10.1111/j.1542-474X.2005.10101.x}.

\bibitem[Thong et~al.(2003)Thong, Li, McNames, Aboy, and Goldstein]{thongAccuracyUltrashortHeart17}
T.~Thong, K.~H. Li, J.~McNames, M.~Aboy, and B.~Goldstein.
\newblock Accuracy of ultra-short heart rate variability measures.
\newblock In \emph{Proceedings of the 25th Annual International Conference of the IEEE Engineering in Medicine and Biology Society}, volume~3, pages 2424--2427, 2003.
\newblock \doi{10.1109/IEMBS.2003.1280405}.

\bibitem[Nussinovitch et~al.(2011)Nussinovitch, Elishkevitz, Katz, Nussinovitch, Segev, Volovitz, and Nussinovitch]{nussinovitchReliabilityUltraShortECG2011}
Udi Nussinovitch, Keren~Politi Elishkevitz, Keren Katz, Moshe Nussinovitch, Shlomo Segev, Benjamin Volovitz, and Naomi Nussinovitch.
\newblock Reliability of {Ultra}‐{Short} {ECG} {Indices} for {Heart} {Rate} {Variability}.
\newblock \emph{Ann Noninvasive Electrocardiol}, 16\penalty0 (2):\penalty0 117--122, 2011.
\newblock \doi{10.1111/j.1542-474x.2011.00417.x}.

\bibitem[Munoz et~al.(2015)Munoz, Van~Roon, Riese, Thio, Oostenbroek, Westrik, De~Geus, Gansevoort, Lefrandt, Nolte, and Snieder]{munozValidityUltraShortRecordings2015}
M.~Loretto Munoz, Arie Van~Roon, Harriëtte Riese, Chris Thio, Emma Oostenbroek, Iris Westrik, Eco J.~C. De~Geus, Ron Gansevoort, Joop Lefrandt, Ilja~M. Nolte, and Harold Snieder.
\newblock Validity of (ultra-)short recordings for heart rate variability measurements.
\newblock \emph{PLOS ONE}, 10\penalty0 (9):\penalty0 e0138921, 2015.
\newblock \doi{10.1371/journal.pone.0138921}.

\bibitem[Burma et~al.(2021)Burma, Graver, Miutz, Macaulay, Copeland, and Smirl]{burma_2021}
Joel~S. Burma, Sarah Graver, Lauren~N. Miutz, Alannah Macaulay, Paige~V. Copeland, and Jonathan~D. Smirl.
\newblock The validity and reliability of ultra-short-term heart rate variability parameters and the influence of physiological covariates.
\newblock \emph{J Appl Physiol}, 130\penalty0 (6):\penalty0 1848--1867, 2021.
\newblock \doi{10.1152/japplphysiol.00955.2020}.

\bibitem[Wu et~al.(2019)Wu, Gu, Yang, and Luo]{wuHowAmusementAnger2019}
Yan Wu, Ruolei Gu, Qiwei Yang, and Yue-jia Luo.
\newblock How do amusement, anger and fear influence heart rate and heart rate variability?
\newblock \emph{Front Neurosci}, 13, 2019.
\newblock \doi{10.3389/fnins.2019.01131}.

\bibitem[Chung et~al.(2025)Chung, Hwang, Kim, Hong, Kim, and Han]{chungChangesHeartRate2025}
Sung~Ah Chung, Hyunchan Hwang, Hee~Jin Kim, Ji~Sun Hong, Sun~Mi Kim, and Doug~Hyun Han.
\newblock Changes in heart rate variability and hemodynamics of adolescents within the frontal cortex in response to face emotional stimulation.
\newblock \emph{PLOS One}, 20\penalty0 (7):\penalty0 e0326204, 2025.
\newblock \doi{10.1371/journal.pone.0326204}.

\end{thebibliography}

\end{document}